\documentclass[12pt]{article}

\usepackage{epsfig}
\usepackage{graphicx}
\usepackage{a4}
\usepackage{latexsym}
\usepackage{cite}

\usepackage{pslatex}
\usepackage[latin1]{inputenc}
\usepackage[T1]{fontenc}

\usepackage{color,colordvi}

\begin{document}

\begin{titlepage}
\begin{flushright}
NIKHEF 07-005\\
SFB/CPP-07-08\\
TTP07-06 \\
{\tt hep-ph/0702279}
\end{flushright}
\vspace{0.8cm}

\begin{center}
\Large
{\bf The Multithreaded version of FORM}
\vspace{1.5cm} \\
\large
M. Tentyukov$^{\, a}$, J.A.M. Vermaseren$^{\, b}$\\
\vspace{1.2cm}
\normalsize
{\it $^a$,Institut f\"ur Theoretische Teilchenphysik \\
\vspace{0.1cm}
Universit\"at Karlsruhe \\
\vspace{0.1cm}
D-76131 Karlsruhe, Germany}\\
\vspace{0.5cm}
{\it $^b$NIKHEF Theory Group \\
\vspace{0.1cm}
Kruislaan 409, 1098 SJ Amsterdam, The Netherlands} \\
\vfill
\large
{\bf Abstract}
\vspace{-0.2cm}
\end{center}
We present TFORM, the version of the symbolic manipulation system FORM that 
can make simultaneous use of several processors in a shared memory 
architecture. The implementation uses Posix threads, also called pthreads, 
and is therefore easily portable between various operating systems. Most 
existing FORM programs will be able to take advantage of the increased 
processing power, without the need for modifications. In some cases some 
minor additions may be needed. For a computer with two processors a typical 
improvement factor in the running time is 1.7 when compared to the 
traditional version of FORM. In the case of computers with 4 processors a 
typical improvement factor in the execution time is slightly above 3.
\vfill
\end{titlepage}
%
%
\setcounter{equation}{0}


\section{Introduction}

The symbolic manipulation system FORM\cite{Vermaseren:2000nd} has been 
available for more than 17 years. It has been used for many calculations 
that involve large intermediate expressions, many of them in quantum field 
theory. It is estimated that the existence of FORM has caused the advance 
of calculations in field theory by one order in perturbation 
theory. None of the more spectacular calculations of refs~\cite{FORMused0,%
FORMused1,FORMused2,FORMused3,FORMused4,FORMused5,FORMused6,FORMused7,%
FORMused8,FORMused9,FORMused10,FORMused11} would have been possible with 
other available systems.

The advantages of FORM over other systems are its capability to handle very 
large expressions and its speed in processing them. Part of the advances in 
the handling of these large expressions can of course be attributed to the 
steady increase in the power of computers. Because the calculations that 
are undertaken require more and more computer power it has been noticed 
that the use of more than one processor at the same time can lead to great 
benefits in a system like FORM. The first attempt took place in 1991 with 
the special computer that had been constructed at FNAL. This experiment was 
however not continued due to problems with access to this computer from 
outside. It led however to a constant awareness of what was needed inside 
FORM to keep the possibility of parallelization and hence its internal 
structure was kept more or less compatible with these needs.

In the late nineties a joint project was started with the University of 
Karlsruhe for adapting FORM to run on special computers with several 
processors. In the beginning this led mostly to a big struggle with the 
hardware and the system software, but when the hardware and the system 
libraries became more stable good results became available. This has led to 
the program ParFORM which uses most of the FORM sources with some extra 
code in addition. It has been described in the literature~\cite{ParFORM1,%
ParFORM2,ParFORM3,ParFORM4,ParFORM5} and several calculations have greatly 
benefited from it. Its main drawback at the moment is that it runs most 
effectively on machines that are rather expensive. Most people however may 
have access to simpler computers with more than a single processor. It is 
nowadays rather common to have computers with two or four processors using 
a shared memory model. It becomes even more relevant because most leading 
vendors switch to dual- and multicore processors. Hence it was judged 
important to create a version of FORM that can make efficient use of such 
systems. Whereas ParFORM uses a message passing protocol named 
MPI\footnote{More about it can be found at http://www.mpi-forum.org/} which 
is quite good for systems that have processors with separated memory or 
consist of networks of computers, computers with a shared memory can work 
with a more efficient communication. The best method here seems the use of 
Posix\footnote{Posix, "a Portable Operating System Interface for uniX" is 
the collective name for a family of related standards.} threads or 
pthreads. A thread is a semi-process, that has its own stack, and executes 
a given piece of code. Unlike a real process, the thread normally shares 
its memory with other threads. Pthreads adhere to strict standards and the 
libraries for them are widely available. These threads can also make more 
efficient use of the available resources as the number of threads is not 
restricted to the number of available processors. Hence if some threads 
have to wait for data, the processors that become available can give 
attention to other threads. The above considerations have led to the 
creation of TFORM, the multithreaded version of FORM, which is described 
below.

When a symbolic system has to operate in a parallel mode one has to worry 
about a number of things. The most important is whether the components of 
an expression can be treated individually. In the case of FORM this can be 
done, because FORM allows only so-called local operations (operations that 
act on only a single term at a time). Another problem might be a central 
variable administration. It should be independent of the order in which the 
terms are processed.

Once the above has been established one has to design strategies for those 
parts of the processing in which information needs to be exchanged between 
the processors. In general there are moments in which expressions need to 
be brought to a unique form. This means that when the contents of an 
expression are spread out over the processors, these processors will have 
to send pieces of the expression to each other. To avoid total chaos there 
should be a single processor that determines which processors will continue 
processing which parts of the expression. Normally this will mean that 
temporarily each expression will be under the complete control of a single 
processor. This constitutes a bottleneck in the processing. In the case of 
FORM this bottleneck consists of two phases: The first is the reading of 
the expression from the disk and distributing the terms over the `workers'. 
The second bottleneck is the final stage of the processing of the terms of 
an expression in which the terms are sorted and put together. This sorting 
is done by merging. In its very last stage it is a single processor that 
will have to do the final merge and write the resulting terms in order to 
the disk. Much attention has to be given to the efficiency of the code for 
these bottlenecks as it puts a limit on the improvement of the combined 
running time.

There are some variables in FORM that are an intermediate between private 
and common variables. Common variables are variables of which there is only 
a single value for every process to use. Private variables are variables of 
which each process has its own copy, or of which each term generates its 
own value. The variables we refer to are called dollar-variables because 
their names start with the dollar character. They can take values and their 
values can be used both during the common phases of a program like 
preprocessing and compilation when only the master process is active, and 
during the execution phases of the program when things are done on a term 
by term basis. In the latter case, when we have a central/common 
administration, their value may depend on the order in which the terms are 
treated. If each processor has its own administration for these variables 
in the end each processor may have a different value, and the question 
arises as to which value we need for the common value. If this problem 
cannot be resolved the corresponding part of the program cannot be 
parallelized. New statements have been implemented to help FORM with such 
conflicts and to minimize the number of cases in which it has to decide to 
run a given module in `sequential' mode. Internally this is done 
differently in ParFORM (each processor has its own administration and the 
resolution of the conflict is at the end of the processing of an 
expression) than in TFORM (one central administration with the resolution 
of conflicts at the moment a variable obtains a value). At the user level 
however this difference shouldn't be noticeable and both will act 
identically.

Special attention has to be given to simultaneous file access. When several 
processors need part of an expression which has been stored on disk, they 
may have to wait for each other. Hence it is best that each processor has 
its own caching system. This will minimize waiting time. Another problem 
occurs in certain systems with the simultaneous use of the disk by the 
processors in the later stages of the sorting of a large expression. We 
have encoutered cases in which the ensuing traffic jams actually made a 
program on 4 processors slower than when it was run on a single processor.

It is rather important that programs need as few modifications as possible 
to be able to use the benefits of running on several processors 
simultaneously. We think we have been able to do this with FORM. This is 
illustrated with a number of test programs that were originally developed 
for running on a single processor (and optimized for it). We try to run 
them unchanged and then measure the improvement in running time (wall clock 
time on a computer that isn't engaged in any other major tasks). The actual 
improvement depends of course on the programming style and particularly on 
what is the ratio of the time spent by the algebraic operations of the 
statements inside the modules versus the sorting at the end of the module. 
The intensive use of external files may influence it also. Finally, work 
done during the compilation and the printing stages is done only inside a 
single processor and hence lowers the efficiency as well.

In the second and third sections we will discuss the issues that needed 
addressing inside the C-code of FORM in order to allow it to run several 
threads simultaneously. In the second section we give attention to the 
modifications of the existing code that were needed before the actual 
parallelization could be attempted. Then in the third section we pass on to 
the FORM specific pieces of code that had to be designed to make the 
multi-threaded running efficient. In the fourth section we test the 
resulting program called TFORM with a number of existing programs that are 
available in the FORM distribution.


\section{Cleaning up the internals of the FORM sources}

The first problem in designing multi-threaded programs is making sure that 
all routines are reentrant or thread-safe. This means that several 
instances of the same routine should be capable of running at the same time 
without unintentionally influencing each other. Hence C-code like
\begin{verbatim}
   static int ScratchArray[100];
   int Multiply(int *term1,int *term2)
   {
   /*
       use ScratchArray to temporarily store pieces of the terms
   */
   }
\end{verbatim}
will usually cause disasters as all instances of the routine Multiply will 
write into the same array at the same time. Therefore we need a very strict 
separation in FORM of which variables are common variables and which 
variables are private variables, with which we mean variables of which each 
thread needs its own copy. Usually this is done by storing all private 
variables in the stack, but in the case of FORM this would involve some 
very big arrays and not all operating systems may be prepared to provide 
that much stack space. In addition it would require much restructuring 
inside FORM. Fortunately the organization of the variables in FORM was 
already such that there was another easy mechanism to make this separation 
and the code didn't have to be modified very much.

All common variables in the sequential version of FORM are part of 
substructures of a single common data structure named A. This was done 
originally to allow compilers to use offsets to the contents of a single 
address register to refer to all common data. At the time it made FORM 
about 10\% faster. These substructures were ordered by type of use, like 
the substructure P containing variables that are used/set by the 
preprocessor, the substructure C containing the variables that concern the 
compiler, the substructure N containing scratch variables used by the 
various routines to communicate with each other during running, etc. Rather 
than referring to these structures as A.P or A.N there are macro's and A.N 
is referred to by the macro AN. All that had to be done was to define a new 
type of data structure and move all substructures (these were the N, R and 
T substructures; all others remained inside the A structure) of which the 
different threads need their own copy to this structure. Each thread will 
allocate its own copy of this structure and we define a new data type B 
which is a pointer to this private structure. After this we redefine the 
macro's AN, AR and AT into B->N, B->R and B->T and once each routine knows 
from which thread it is running and hence knows the value it needs for this 
variable B, the rest of the routine needs no further changes. Because all 
parameter fields to the routines are already run by macro's (originally 
this was to facilitate the differences between ANSI-C and non-ANSI-C), it 
was also easy to redefine some macro's and automatically pass the pointer B 
to routines that need it.

The above changes affect nearly all the code and make the program slightly 
slower on average although we have met rare cases in which TFORM on a 
single processor is a little bit faster than the sequential version with 
its original definitions of the macro's.

The next changes concern the reading of files. In the sequential version it 
was possible to position a file in one routine and read it in another. One 
could also assume that a file was still at the position at which it was 
left after the previous access. In a parallel environment this is no longer 
possible. Hence one needs to use locks (for which one can use so-called 
mutex variables). But because one wants to lock the access to a file for as 
short a time as possible, it is important to have the positioning of the 
file and the access to it as close together as possible. This required a 
number of modifications and was a rather annoying source of errors. The 
caching system had to be adapted as well to allow each processor to have 
its own cache, even though the file is opened only once. The way the system 
works now is that if an expression fits completely inside memory and hence 
lies completely inside the cache\footnote{The size of such a cache is 
determined by the setup variable ScratchSize.} of the master processor, all 
worker processors read from the cache of the master, but they have their 
own set of pointers to from where in the cache they are reading. If the 
expression is too big for the cache of the master and is actually residing 
on disk, all workers will use their own cache of which the size is $1/N$ 
times the size of the cache of the master with $N$ the number of worker 
threads. These worker caches are in addition to the cache of the master and 
hence in total we need twice the size of the cache of the master. It is 
hard to improve on this as the master also needs the cache for reading the 
input and distributing the terms over the workers. Trying to improve on 
this would add much complexity to the code.

The next file problem comes with the stored expressions. It is somewhat 
related to the relocation problem when the executable code of a program is 
being loaded into the memory, just before running. In the sequential 
version expressions are stored with their symbol table. This may tell that 
for instance the variable x is represented by the symbol number 5 inside 
the stored expression. When the stored expression is used, its symbol table 
is compared with the table of active objects and the variable x may either 
not be in it yet or may have a different number. If necessary, the variable 
x will be entered in the symbol tables. In any case, a renumbering table 
has to be made because the symbol number of x may now be different, for 
instance 12 rather than the 5 it has in the stored version. Originally this 
would be done during the execution of the program at the moment an 
expression would be needed. In a parallel environment however different 
terms may use different expressions. This means that the variables may not 
be added in the same order when several processors are trying to do this 
simultaneously. Hence it should be clear that when several threads want to 
do this at the same time we need either a sophisticated system of locks 
around the common variable administration, or we can make errors. The 
solution selected was to let the compiler (which runs on the single master 
thread) read the symbol tables of the expressions that {\sl might} be used. 
This avoids conflicts in the numbering without having to resort to locks. 
On the other hand, if an expression is not used because during running the 
control flow avoids the statement in which it would be used, its symbols 
are read nevertheless. This could potentially result in wasteful use of 
space in the name tables. It was however considered the lesser evil.

For stored expressions multiple caching is also very important. A new 
caching system was implemented of which the sequential version of FORM also 
benefits. Therefore one may notice that, also in the sequential version of 
FORM, programs that make intensive use of stored expressions will execute 
faster now.

The final cleanup of the code concerned the pattern matching where use was 
made of some common variables in a very intransparent way. These had to be 
transferred to the private data space.

The above cleanup concerns all versions of FORM. Hence the moving of the 
positioning statements of files have their effect on the sequential version 
as well. By working this way it is much easier to maintain FORM as 
basically there is only a single version of the source code. The setting of 
just one parameter controls whether the compiler will produce the 
sequential version or the multithreaded version.


\section{The parallelization}

For the parallelization we use the conventions of the Posix threads, 
shortly called pthreads. These are implemented by means of the pthread 
library, which is available on all modern UNIX systems. For the definition 
of the concepts and their use we used ref.\cite{pthreads}. The model we use 
consists of the startup of one master thread and $N$ worker threads in a 
so-called thread pool. This seems most efficient on a computer with $N$ 
processors. Often when the master has to do work, at least one of the 
workers is idle and conversely when all workers are occupied, often the 
master has very little to do. We will show in the tests that indeed this 
works well.

The workers are started only once at the startup of TFORM. They allocate 
their own private memory inside a single data structure of which the address 
(called B as explained above) is stored in a common array from which the 
address can be recovered when the number of the worker is known. On 
computers on which the threads are moved between the processors in a 
semi-random fashion, the individual allocations, rather than a single 
common allocation that is parcelled up afterwards, leads to no benefits, 
but it allows to create a coherency between workers and memory on those 
computers for which it does make a difference. Currently nothing is done 
with this.

Next all private arrays for caching and sorting are allocated by the workers 
and the master. After this the workers go to sleep. The master waits till 
all workers are sleeping and then starts the execution of the FORM program 
in which the preprocessor and the compiler treat each module before 
executing it. The execution phase of each module is the part that is 
eligible for parallelization.

In FORM each expression is executed in sequence which means one after the 
other. The same happens with the terms inside the expressions. The order is 
determined by the final sorting in the previous module. In TFORM (as well 
as in ParFORM) the master process reads the terms of one expression and 
distributes them over the workers. The terms are usually bunched together 
in something called buckets. This is because signals\footnote{Technically 
they aren't signals. Calls to pthread\_cond\_signal() are part of the 
Pthreads mechanisms for synchronizing processes. For the sake of simplicity 
we will refer to them as signals.} have to be sent to the worker that 
should receive the terms and these signals turn out to be rather costly. 
For many programs the optimal bucket size lies between 100 and 1000 terms. 
Currently the default has been set to 500 terms. To keep the workers 
waiting as short as possible, the total number of buckets is twice the 
number of workers. This way the master process can prepare spare buckets. 
Once a worker becomes available all the master has to do is to copy some 
necessary `environment' variables to the private data structure of the 
worker and the pointer to the bucket. After that a wakeup signal can be 
sent.

The next critical point is when the last terms in the input expression have 
been sent. After this workers will become available, but there are no 
further terms for them. This can become inefficient, if there is a single 
worker that has one or more terms that require much CPU time. Often such 
difficult terms tend to be grouped together and hence they could be inside 
the same bucket. For this a load balancing system has been designed in 
which the master processor will look inside the buckets of the workers and 
possibly steal some terms back to give to idle workers. This works well, 
but it fails in the case that there is a single term that uses most of the 
CPU time. Such is the case in the following FORM code:
\begin{verbatim}
    Symbols x1,...,x10;
    Local F = (x1+...+x10)^10;
    id  x10 = 1-x1-...-x9;
    .end
\end{verbatim}
A new expressions always starts as a single term and inside the module it 
will be expanded. Hence the above example cannot use the parallelization, 
because its single term will end up inside a single worker.

In principle FORM is designed in such a way that also this load can be 
balanced. It is possible to interfere in the expansion tree and organize 
the delegation of subtasks to idle co-workers. It does however need a 
number of strategic decisions about when and where in the tree this 
interference should take place. This will be considered in a later version. 
For now one should remember that if one would like to make efficient use of 
the parallelization, one should avoid having the early modules doing too 
much work and keep the work for the modules in which there are already a 
large number of terms at the input. In the case of the above example one 
could benefit from parallelization if the program would read
\begin{verbatim}
    Symbols x1,...,x10;
    Local F = (x1+...+x10)^10;
    .sort
    id  x10 = 1-x1-...-x9;
    .end
\end{verbatim}
because the id-statement which does most of the work, is now executed at a 
moment that the master can distribute terms over the workers. At the same 
time the overhead of the extra sort is still negligible compared to the 
total amount of work to be done.

Once the input terms have all been processed, it is time to do the final 
sorting. Here we are faced with the second bottleneck, because now the 
completed expression should be put together under the control of the single 
master processor. The idea is to have the workers do as much of the sorting 
as possible. The master will then merge the sorted results of the workers.

We tried several methods for this sorting. In the first version of TFORM 
each worker did a regular sort as is done in the sequential version of 
FORM. This means that each might use a sort file and each would write its 
output to an output scratch file. Then the master would merge these files 
in the same way that the patches in a single sort file would be merged. The 
only difference was that rather than reading from different sorted patches 
inside a single file, the patches would come from different files. The 
results were disastrous. Apparently the LINUX file system that was used 
(ReiserFS) becomes very inefficient when several files are accessed at the 
same time by the same program. The test program running on four processors 
made the program actually take more than four times as much real time as 
when it was running on just a single processor, and the whole computer 
became very slow. The exact cause is still not clear and the issue remains 
under investigation.

The solution that has been selected cuts out the output scratch files for 
the workers. Instead the workers write their output directly into the sort 
buffers of the master. The master will then sort them simultaneously, 
provided this is possible. A system of data blocks has been set up to allow 
this simultaneous work. Each `stream' from the workers inside the sort 
buffer of the master is divided into 10 blocks that are arranged in a 
circular fashion. Locks prevent the master from accessing blocks that have 
not been filled by the workers and workers from writing into blocks that 
have not yet been treated by the master. The result is that now each worker 
needs only one file, its sort file. Performance is much better, but still 
there are problems with the LINUX file system we used. The same program 
that needed four times as much real time in the first approach, now needs a 
bit more than half the amount of real time than the sequential version. And 
during some of the sorting the computer is virtually unusable from the 
terminal. This depends very much on the way the file activity is organized. 
The size of the sorting buffers and the cache are of great influence. It 
also depends on which computer is used. We illustrate this with the first 
example in the next section (case N=15). In all other examples however this 
effect doesn't occur.

The next problem to address is the treatment of the dollar variables. These 
variables are in principle common but if one isn't carefull each processor 
may obtain a private value and hence a they are a 
potential source of conflicts. Let us have a look at some examples:
\begin{verbatim}
   #$m1 = 0;
   #$mc = 0;
   if ( count(x1,1) > $m1 ) $m1 = count_(x1,1);
   if ( count(y,1) > 0 ) $mc = $mc+1;
   if ( match(f(x?$mx)) );
      id g($mx,n?) = n+$mx;
   endif;
   .sort
\end{verbatim}
We have here three uses of dollar variables that can lead to problems. In 
the case of the variable \verb:$m1: two workers could make the compare at 
the same time, after which the eventual value might not become the maximum 
of those two. In the case of \verb:$mc: one could have that worker 1 picks 
up the value, let us say 6, then worker 2 also picks up 6, then worker 1 
writes 7, after which worker 2 writes 7. And similar accidents can happen 
with \verb:$mx:.

In the case of these accidents the placing of locks is counterproductive. 
One would have to place the lock around most of the substitution tree and 
hence the workers would obstruct each other rather thoroughly. There is 
however a solution. If one can tell TFORM what is meant with the dollar 
variable, TFORM might be able to deal with it without much overhead. This 
is done with the ModuleOption statement at the end of the module. In this 
case we would add the statement
\begin{verbatim}
   ModuleOption,maximum,$m1,sum,$mc,local,$mx;
\end{verbatim}
before the .sort instruction. This way TFORM knows that we collect a 
maximum in \verb:$m1: and hence makes a compare, if needed locks the 
variable, makes a new compare (!) and if needed stores the new value. Then 
it releases the lock on the variable. This implementation has a side 
effect. The following code would produce the same result:
\begin{verbatim}
    #$m1 = 0;
    $m1 = count_(x1,1);
    ModuleOption,maximum,$m1;
    .sort
\end{verbatim}
The maximum declaration in the ModuleOption statement makes the original if 
statement superfluous. It would however be very unwise to program this way 
because the sequential version of FORM just ignores this part of the 
ModuleOption statement and hence would not produce the proper result. Also 
the different working of ParFORM would not give the correct result.

The case of the sum is more complicated. We have to place a lock 
immediately when the statement starts its evaluation and we can release the 
lock only after the new assignment. This is rather expensive and therefore 
should be used with great care. Of course this sum option can be abused. 
One could apply it to other assignments that are not sums. One is strongly 
advised against this, because in the future the inner workings may change 
and only a proper sum will work correctly\footnote{One could think of 
substituting zero for the occurrence of \$mc in the right hand side and 
adding the obtained value for the complete right hand side into the 
existing value of \$mc. This would reduce the time for locking the variable 
by a considerable amount.}. In the case of a local use of a dollar variable 
(like \verb:$mx:), each thread will have its own copy. If there was a 
common value at the start of the module, this private/local variable will 
be initialized by it. At the end of the module the private copy is deleted. 
The common copy will still have the value that it had at the start of the 
module.

In ParFORM the treatment of the above dollar variables is somewhat 
different. Each processor has its own copy of the variable administration. 
Hence \verb:$m1: will become a local/private maximum and \verb:$mc: a sum 
over all terms treated in that processor. Only at the end of the module the 
private values are combined into a common value, using the information in 
the ModuleOption statement.

Of course there can be uses of dollar variables that cannot be 
parallelized. If TFORM is not helped in its use of dollar variables it will 
automatically switch to sequential mode for the module(s) in which this 
dollar variable is given a value during execution. Hence old programs will 
continue to run. They may however not benefit from the parallelization in 
such modules. The sequential version of FORM accepts the ModuleOption 
statement, but it just ignores most of its options. When writing programs 
that use dollar variables, it is best to plan the ModuleOption statements 
with it. This way the program will run optimally both in the sequential and 
in the parallel mode.

The only other instance of the use of local/private definitions of common 
variables concerns the preprocessor variables as redefined by the redefine 
statement. At first one might think that here we run into the same problems 
as with the dollar variables, but this is not the case. The value of a 
preprocessor variable can only be used by the preprocessor. Therefore 
during execution it is a write-only variable. If we remember which term was 
responsible for the last redefinition, we have to make the redefinition 
only when the number of the current input term is greater or equal to the 
number of the input term that caused the last redefinition. Hence this 
problem has been solved inside FORM and needs no special attention from the 
user.


\section{The performance}

The above modifications allow the parallel running of all existing FORM 
programs. The only exception are the modules in which dollar variables are 
assigned during execution time and that are not aided by ModuleOption 
statements. These will be run in sequential mode. The running itself is 
rather simple. One calls TFORM in the same way FORM used to be called but 
an extra argument in the command tail is needed to specify the number of 
worker threads as in:
\begin{verbatim}
   tform -w4 -l calcdia
\end{verbatim}
The -w argument specifies the number of workers. If this number is zero or 
one, there will only be a master thread. Invocation of a single worker 
would only cost overhead for communication and wouldn't improve 
performance. Omission of the -w argument is the same as -w0.

The first testcase we use is a simple program for computing so-called 
chromatic polynomials. The program is described in a set of example files 
in the FORM distribution. We use the program named p15.frm for a variety of 
lattice sizes. The main part of the code is a loop given by
\begin{verbatim}
  Multiply F`i';
  repeat id  d(?a,k?,?b)*d(?c,k?,?d) = d(?a,?c,k,?b,?d);
  Symmetrize d;
  repeat id  d(?a,k?,k?,?b) = d(?a,k,?b);
  #do d = 1,`D'
    id  d(k`i',?a,k1?kk0[x],?b,k2?kk`d'[x],?c) = 0;
  #enddo
  id,ifmatch->1,d(k`i',k?) = 1;
  id,ifmatch->1,d(k`i',?b) = d(?b);
    Multiply acc([q-1]+1);
  Label 1;
  id  d = 1;
  .sort:`i';
\end{verbatim}
The loop contains a few id-statements with non-trivial pattern matching and 
some polynomial arithmetic inside the `polyfun' named acc. We ran this 
program on a variety of computers.


The first computer, designated P, has two Pentium4 processors at 1.7 GHz. 
The second computer, designated N, contains 4 Opteron processors at 2.6 
GHz. The third computer is a SGI computer with 32 Itanium processors at 1.3 
GHz. Although the last computer uses shared memory there is a 
hierarchy. Also one has to specify how many and which processors are 
assigned to the job. We will designate the SGI computer S\# in which \# is 
the number of processors reserved for the task. These processors are 
usually selected to have the best memory latency for the selected number. 
In one case we did a test in which we selected the memory in such a way 
that the situation was the worst possible. This made an almost two percent 
difference in execution time. Hence the effect doesn't seem very relevant. 
In the case that we specify zero workers we used the sequential version of 
FORM. In the case we specify one worker we used TFORM with only the master 
thread.
\begin{center}
\begin{tabular}{|c|c|c|r|r|} \hline
	Lattice & Computer & Workers & Time(sec)& Improvement \\ \hline
       10   &    N     &    0    &  12.45   &             \\
       10   &    N     &    1    &  11.92   &   1.0445    \\
       10   &    N     &    2    &   6.64   &   1.8750    \\
       10   &    N     &    3    &   4.83   &   2.5776    \\
       10   &    N     &    4    &   4.30   &   2.8953    \\
       10   &    N     &    5    &   4.37   &   2.8490    \\
       10   &    N     &    6    &   4.35   &   2.8621    \\
       10   &    N     &    7    &   4.30   &   2.8953    \\
       10   &    N     &    8    &   4.36   &   2.8555    \\
       10   &    N     &   16    &   4.50   &   2.7667    \\
       10   &    N     &   32    &   4.95   &   2.5152    \\ \hline
       10   &    P     &    0    &  29.96   &             \\
       10   &    P     &    1    &  30.67   &   0.9769    \\
       10   &    P     &    2    &  17.45   &   1.7169    \\ \hline
\end{tabular} \vspace{3mm} \\
Table 1: Performance for chromatic polynomials on simple lattices.
\end{center} \vspace{0.5cm}
We notice here a number of things. One is that on the Opteron this program 
is a rare example in which TFORM in single thread mode is faster than the 
sequential version of FORM. Next we see that there is a saturation effect, 
because the computer N has four processors. With more than 4 workers the 
improvement factor starts going down again. At first the effect is not very 
strong. In the case of 32 workers though the master has to start doing more 
work in the sorting as it has to merge 32 streams which gives on average 5 
compares per term. This effect becomes noticeable. The optimum is at 4 
workers, as the master seems to be able to get enough time on the various 
processors. In total the master processor used 1.16s of CPU time in that 
run. Finally we make a remark about running a smaller number of threads 
than there are processors. In the case of two workers on a machine with 
four processors the master can run simultaneously with the workers. This 
can make a difference during the final stages of the sorting and during the 
filling of the input buckets. Hence the efficiency will be slightly higher 
than it would be on a similar machine with two processors. We see this 
effect here when we compare the efficiencies for two workers on the 
machines N (4 processors) and P (2 processors), although it is not 
completely clear whether this is the only reason for the difference.

For bigger lattices we obtain:
\begin{center}
\begin{tabular}{|c|c|c|r|r|} \hline
	Lattice & Computer & Workers & Time(sec)  & Improvement \\ \hline
       14   &    N     &    0    &  2809.57   &             \\
       14   &    N     &    1    &  2695.64   &   1.0423    \\
       14   &    N     &    2    &  1443.07   &   1.9469    \\
       14   &    N     &    3    &  1059.07   &   2.6529    \\
       14   &    N     &    4    &   923.86   &   3.0411    \\ \hline
       14   &    P     &    0    & 15786.37   &             \\
       14   &    P     &    2    &  9163.65   &   1.7227    \\ \hline
       15   &    N     &    0    & 12541.39   &             \\
       15   &    N     &    1    & 12234.77   &   1.0251    \\
       15   &    N     &    2    &  7601.00   &   1.6500    \\
       15   &    N     &    3    &  5332.38   &   2.3519    \\
       15   &    N     &    4    &  3892.80   &   3.2217    \\
       15   &    N     &    6    &  4046.58   &   3.0993    \\
       15   &    N     &    8    &  4100.53   &   3.0585    \\ \hline
       15   &    P     &    0    & 77349.12   &             \\
       15   &    P     &    2    & 45996.63   &   1.6816    \\ \hline
       15   &   S0     &    0    & 21897.79   &             \\
       15   &   S2     &    2    & 13853.82   &   1.5806    \\
       15   &   S4     &    4    &  7867.53   &   2.7833    \\
       15   &   S8     &    8    &  6622.27   &   3.3067    \\
       15   &   S16    &   16    &  5543.48   &   3.9502    \\
       15   &   S32    &   32    &  5664.99   &   3.8655    \\ \hline
\end{tabular} \vspace{3mm} \\
Table 2: Performance for chromatic polynomials on bigger lattices.
\end{center} \vspace{0.5cm}
In the case of the 14x14 lattice on the computer N the expression could 
still fit inside the allocated buffers. We see a similar pattern as before. 
For the 15x15 case however the sorting stage needed heavy use of disk 
files. We used a large sorting buffer of 2 Gigabytes. This is still not 
large enough to avoid the use of a sort file for each of the workers. One 
can see that the behaviour is much better on the Opteron computer (N) than 
on the SGI machine. To compare we also ran with smaller buffers (400 
Mbytes). This causes many more and smaller disk operations. It gives the 
following results:
\begin{center}
\begin{tabular}{|c|c|c|r|r|} \hline
	Lattice & Computer & Workers & Time(sec)  & Improvement \\ \hline
       15   &    N     &    0    &  14807.98  &             \\
       15   &    N     &    2    &   9834.92  &   1.5057    \\
       15   &    N     &    4    &   7833.86  &   1.8903    \\ \hline
       15   &   S0     &    0    &  25312.04  &             \\
       15   &   S2     &    2    &  15208.59  &   1.6643    \\
       15   &   S4     &    4    &  12879.12  &   1.9654    \\
       15   &   S8     &    8    &  11832.35  &   2.1392    \\
       15   &   S16    &   16    &  10490.21  &   2.4129    \\
       15   &   S32    &   32    &  10200.72  &   2.4814    \\ \hline
\end{tabular} \vspace{3mm} \\
Table 3: As in table 2 but with smaller buffers.
\end{center} \vspace{0.5cm}
Clearly lots of (nearly) simultaneous disk operations have a bad effect on 
the performance. It is guessed that this effect is due to the particular 
type of file system used (ReiserFS) as another file system doesn't show 
this effect\footnote{This was tested with heavy file copy operations rather 
than with TFORM.} The slower computer P, which has less memory and hence 
smaller buffers anyway, has already problems with the disk when running in 
the sequential mode and hence the effect seems to be less pronounced.  The 
different ratio in the speed between the computers as compared to the 
$N=10$ lattice is due to several factors. First the bigger coefficients in 
the bigger lattices make the 64 bits Opteron more efficient. And second the 
Opteron computer has much more memory. For the determination of the 
improvement factor this is however irrelevant.


The second example concerns a number of Feynman diagrams as computed for 
past publications. The program used for these computations is called mincer 
and it has been heavily optimized for use with version 2 of FORM. Later 
some extra optimizations were added for the use with version 3. Not a 
single line was changed for the runs with TFORM. We label the diagrams d1c, 
d10c and d11c as they come from a set computed for the nonsinglet sector of 
the form factor F2 in deep inelastic scattering. They are three loop 
diagrams of the non-planar type and we calculate Mellin moments. 
\begin{center}
\begin{tabular}{|c|c|c|c|r|r|} \hline
  Diagram & Moment & Computer & Workers & Time(sec)& Improvement \\ \hline
    d1c   &   10   &    N     &    0    &   32.70  &             \\
    d1c   &   10   &    N     &    1    &   33.41  &   0.9799    \\
    d1c   &   10   &    N     &    2    &   18.29  &   1.7900    \\
    d1c   &   10   &    N     &    3    &   12.85  &   2.5479    \\
    d1c   &   10   &    N     &    4    &   10.39  &   3.1511    \\ \hline
    d1c   &   16   &    N     &    0    &  694.02  &             \\
    d1c   &   16   &    N     &    1    &  727.52  &   0.9540    \\
    d1c   &   16   &    N     &    2    &  387.74  &   1.7899    \\
    d1c   &   16   &    N     &    3    &  266.23  &   2.6068    \\
    d1c   &   16   &    N     &    4    &  208.74  &   3.3248    \\ \hline
    d1c   &   10   &    P     &    0    &  108.55  &             \\
    d1c   &   10   &    P     &    2    &   60.32  &   1.7996    \\
    d1c   &   16   &    P     &    0    & 2759.26  &             \\
    d1c   &   16   &    P     &    2    & 1463.70  &   1.8851    \\ \hline
    d10c  &   10   &    N     &    0    &  880.66  &             \\
    d10c  &   10   &    N     &    1    &  894.69  &   0.9843    \\
    d10c  &   10   &    N     &    2    &  474.23  &   1.8570    \\
    d10c  &   10   &    N     &    3    &  328.68  &   2.6794    \\
    d10c  &   10   &    N     &    4    &  263.33  &   3.3443    \\ \hline
    d11c  &   10   &    N     &    0    & 3208.02  &             \\
    d11c  &   10   &    N     &    1    & 3271.74  &   0.9805    \\
    d11c  &   10   &    N     &    2    & 1719.05  &   1.8662    \\
    d11c  &   10   &    N     &    3    & 1221.12  &   2.6271    \\
    d11c  &   10   &    N     &    4    &  971.87  &   3.3009    \\ \hline
\end{tabular} \vspace{3mm} \\
Table 4: Runs with Mincer for three different diagrams.
\end{center}
The diagram d11c was the `most difficult' diagram in the set of all 
diagrams that had to be computed for this reaction. Because more work is 
done inside the modules the efficiency here is higher than in the previous 
example. One has to consider that in the case of multithreaded runs there 
is an overhead of 5\% - 10\% as compared to the sequential version. This 
means that the theoretical limit is an improvement factor of about 3.6. on 
a computer with 4 processors. One has to add that there is also a part of 
the time in which only the master thread is active. Hence an improvement 
factor of 3.3 on a machine with four processors is quite high.
\begin{center}  
\begin{tabular}{|c|c|c|c|r|r|r|} \hline
  Diagram & Moment & Computer & Workers & Time(sec)& Master   & Improvement \\ \hline
    d11c  &   10   &    S1    &    0    & 7042.83  &          &             \\
    d11c  &   10   &    S2    &    2    & 3834.41  &  196.21  &   1.8367    \\
    d11c  &   10   &    S3    &    3    & 2625.51  &  273.05  &   2.6825    \\
    d11c  &   10   &    S4    &    4    & 2117.88  &  291.37  &   3.3254    \\
    d11c  &   10   &    S6    &    6    & 1644.43  &  392.37  &   4.2828    \\
    d11c  &   10   &    S8    &    8    & 1335.42  &  418.30  &   5.2739    \\
    d11c  &   10   &    S12   &    12   & 1093.35  &  483.02  &   6.4415    \\
    d11c  &   10   &    S16   &    16   & 1002.00  &  522.96  &   7.0288    \\
    d11c  &   10   &    S24   &    24   &  953.70  &  606.45  &   7.3847    \\
    d11c  &   10   &    S32   &    32   &  957.79  &  658.27  &   7.3532    \\ \hline
\end{tabular} \vspace{3mm} \\
Table 5: Runs with Mincer on the SGI computer using TFORM.
\end{center}
On the sgi machine we notice a clear example of saturation. Till 4 threads 
the improvement is quite nice. From 4 to 16 threads the improvement isn't 
very spectacular but might still be worth it. Above that one can hardly see 
any further improvement and one is just wasting resources. For 32 
processors the improvement even worsens. The dominant reason is clearly the 
time spent by the master process. This increases when more threads are 
involved. The two effects one can think of immediately are the extra work 
that has to be done by the master in the final sorting and the fact that 
more threads have to be provided with data, running the risk that threads 
have to wait. This last effect seems to be present, but even if it wouldn't 
be present, the efficiency would be no more than about 8 for 32 processors. 
If we ignore this last effect (which can be controled somewhat with the 
setting of the variable ThreadBucketSize) the real time would roughly 
follow a formula like
\begin{eqnarray}
	T & = & c_1 + c_2/N + c_3\ {}^2\log N
\end{eqnarray}
in which N is the number of workers and $c_1,c_2,c_3$ are constants.
One can read from the table that each extra compare that has to be done by 
the master thread (5 in the case of 32 processors) adds more than 100 sec 
to its CPU time and hence $c_3 \approx 115$. Therefore the sorting seems to 
be the main culprit. Clearly much of the saturation can be alleviated by a 
future change in the workload of the master during the final stages of the 
sorting. As the values of $c_1,c_2,c_3$ depend on the problem the exact 
value of the saturation in the improvement factor is hard to predict.

We managed to compare this last example with the current version of 
ParFORM. This gives similar results as can be seen from the next table:
\begin{center}  
\begin{tabular}{|c|c|c|c|r|r|} \hline
  Diagram & Moment & Computer & Workers & Time(sec)& Improvement \\ \hline
    d11c  &   10   &    S1    &    0    & 6998.00  &             \\
    d11c  &   10   &    S3    &    2    & 3704.21  &   1.8892    \\
    d11c  &   10   &    S4    &    3    & 3353.51  &   2.0868    \\
    d11c  &   10   &    S5    &    4    & 2117.88  &   3.3042    \\
    d11c  &   10   &    S8    &    7    & 1642.14  &   4.2615    \\
    d11c  &   10   &    S9    &    8    & 1374.69  &   5.0906    \\
    d11c  &   10   &    S13   &    12   & 1221.45  &   5.7293    \\
    d11c  &   10   &    S16   &    15   & 1163.73  &   6.0134    \\
    d11c  &   10   &    S17   &    16   & 1088.27  &   6.4304    \\
    d11c  &   10   &    S24   &    23   & 1074.26  &   6.5143    \\
    d11c  &   10   &    S25   &    24   &  986.38  &   7.0946    \\
    d11c  &   10   &    S32   &    31   & 1010.73  &   6.9237    \\ \hline
\end{tabular} \vspace{3mm} \\
Table 6: Runs with Mincer on the SGI computer using ParFORM.
\end{center}
We see here the same saturation. One should also realize that in ParFORM 
one processor is reserved for the master process exclusively. This explains 
the rather bumpy transition when the number of processors is around a power 
of two. On the whole, the slower MPI communication costs a marginal amount 
of efficiency. This situation is better when the size of the various 
buffers is tuned to ParFORM and the specific computer. In that 
case S32 reaches an efficiency of 7.4247. In the case of an experimental 
version of ParFORM in data is transfered via shared memory the maximum 
efficiency becomes 7.7132 for S32. 


The final example concerns the solution of a large system of equations. The 
equations in question form all relations that are known between multiple 
zeta values of the same weight. This system was described in 
ref~\cite{Vermaseren:1998uu} and the programs can be found in the FORM 
distribution under `summer'. In ref~\cite{Vermaseren:1998uu} the equations 
were only worked out and solved till weight 9, but because the computers 
have become faster and we have parallel processing we decided to try weight 
10 as well. It turned out that most of the CPU time for the weight 10 
calculations is used for the calculation of GCD's of large integers (like 
300 decimal digits). Considering that originally FORM was optimized for 
relatively short integers and that hence some improvements might be 
possible, the GCD algorithms were studied carefully and indeed a 
significant improvement for large integers was found. This made the runs of 
the weight 10 programs faster by a factor 6.95. For weight 9 the 
improvement was a factor 2.76 as there the numbers were not quite as large. 
We will give the timings with the new version\footnote{Because the 
calculations of the GCD's is mostly local/private during term generation and 
normalization, the efficiency of the parallelization is somewhat better for 
the old version. We have reached there a record factor of 3.49 for the 
weight 10 calculation on 4 processors, but the total execution time was of 
course much larger than with the new version.}. It should be noted that a 
part of the program in which a number of equations are prepared for a 
Gaussian elimination is better done inside a single processor. This does 
spoil the parallelization a bit for the lower weights. For the higher 
weights most work is in the rational arithmetic, even with the improved 
routines, and hence this gives a much better efficiency. But it does give a 
relatively large value for the CPU time used by the master process.
\begin{center}
\begin{tabular}{|c|c|c|r|r|r|} \hline
  Weight & Computer & Workers &   Time(sec)& CPU master(sec)& Improvement \\ \hline
     7   &    N     &    0    &      8.43  &             &             \\
     7   &    N     &    4    &      8.74  &      4.91   &   0.9645    \\
     8   &    N     &    0    &     78.71  &             &             \\
     8   &    N     &    4    &     62.73  &     36.92   &   1.2547    \\
     9   &    N     &    0    &   2026.15  &             &             \\
     9   &    N     &    2    &   1330.41  &    372.65   &   1.5230    \\
     9   &    N     &    4    &    909.47  &    373.40   &   2.2278    \\
    10   &    N     &    0    & 130860.59  &             &             \\
    10   &    N     &    2    &  71479.05  &   5510.02   &   1.8308    \\
    10   &    N     &    4    &  39151.96  &   5590.78   &   3.3424    \\ \hline
    10   &   S1     &    0    & 233797.28  &             &             \\
    10   &   S2     &    2    & 125072.98  &  10013.77   &   1.8693    \\
    10   &   S4     &    4    &  69075.57  &  10645.14   &   3.3847    \\
    10   &   S8     &    8    &  41987.38  &  11693.31   &   5.5683    \\
    10   &  S12     &   12    &  32910.15  &  12194.61   &   7.1041    \\
    10   &  S16     &   16    &  28317.13  &  12564.38   &   8.2564    \\
    10   &  S24     &   24    &  24108.43  &  13189.96   &   9.6977    \\
    10   &  S32     &   32    &  22143.32  &  13633.52   &  10.5584    \\ \hline
\end{tabular} \vspace{3mm} \\
Table 7: Solving dependencies between multiple zeta values.
\end{center}
Beyond weight 9 the test couldn't be done on 32-bits architectures, because 
the size of the tables would be too large (FORM has a limitation on the 
number of elements in a single table which depends on the word size). We 
see on the Silicon Graphics computer that the CPU time of the master 
process becomes the limiting factor in the efficiency when we increase the 
number of processors. The eventual value of the improvement factor is of 
course a function of how much work can be done completely locally. As the 
rational arithmetic is very local, the factor here is somewhat better than 
in the previous examples. The waiting of workers for input terms and the 
signals involved in sending them were quite a factor here. The runs we show 
here were with a ThreadBucketSize of 1000. With a value of 100 the 
improvement factor for 32 processors went down to 7.75.

It is also clear that when the equations are relatively simple the solution 
selected doesn't benefit much from parallelization. This is in the nature 
of the problem, and only because of the vast amount of rational arithmetic 
the more complicated cases give good improvements.


\section{Conclusions}

TFORM is working and can handle in principle all existing programs that 
would run on version 3.1 of FORM\footnote{It is of course not excluded that 
the extensive changes have caused the inclusion of some new bugs that still 
have not been caught. If encountered, please report them to the authors.}.
On computers with a limited number of processors the improvement in running 
time is quite good. The more complicated the program the better the 
improvement.

We have compared the performance of TFORM with that of ParFORM. As is to be 
expected, when the two can run on the same computers they give more or less 
the same increase in efficiency. Each have their own strong points. ParFORM 
can run on a larger number of multi processor systems. The internal 
organization allows TFORM to run more existing programs in parallel. Also 
for a small number of processors TFORM can run more efficiently as it 
doesn't have to reserve a whole processor for the master.

For computers with a larger number of processors there is a strong 
saturation effect due to the increase in tasks that have to be done by the 
master thread. The way to improve TFORM in the future has to be sought in 
the lightening of the load of the master process. One way is to remove the 
signals that are sent around between the master and the workers when the 
terms are distributed over the workers. The most important however seems to 
be a change in the final stages of the sorting. In principle it is possible 
to set this up as a binary tree in which all but the last step are done by 
workers, and maybe it is even possible to let a worker do the last step, 
making the master only responsible for writing the final result. This will 
be tried in a future version of TFORM. There will however always be a 
program dependent limit as a sorting tree will always imply a bottleneck in 
which one compare per term is to be done inside a single 
thread\footnote{One could imagine a sorting system in which two processes 
are working simultaneously, one from the smallest term upward and one from 
the largest term downward. This creates an enormous complexity while the 
benefits are relatively limited.}. In the ideal case the execution time 
will follow a rule like
\begin{eqnarray}
	T & = & c_1 + c_2/N
\end{eqnarray}
in which the values of $c_1$ and $c_2$ are problem dependent. One can 
try to optimize programs in such a way that for a given number of workers 
the execution time is optimal. We have not yet experimented with this but 
we expect that some users will do this in the future.

Considering the bottlenecks it should be clear that at the moment only for 
special problems the use of a very large number of processors can be 
beneficial. For a small number of processors however the current version of 
TFORM is more than adequate.

Acknowledgements: The work of M. Tentyukov is supported by the DFG, 
SFB-TR9. The work of J. Vermaseren is supported by FOM. The authors thank 
Y. Schr\"oder for taking the time to prepare a nontrivial bug report.



\end{document}